# IoT-Enabled Sleep Monitoring and Cognitive Assessment for Evaluating Teacher Well-Being

Anwar Ahmed Khan
*Department of Computer Science,*
*Millennium Institute of Technology and Entrepreneurship*
Karachi, Pakistan
yrawna@yahoo.com
0000-0002-2237-5124

Shama Siddiqui
*Department of Computer Science,*
*DHA Suffa University*
Karachi, Pakistan
shamasid@hotmail.com
0000-0002-3547-0307

Mehar Ullah
*Department of Electrical Engineering,*
*LUT University,*
Lappeenranta, Finland
mehar.ullah@lut.fi
0000-0003-2405-8998

Indrakshi Dey
*Walton Institute for Information and Communications Systems Science.*
*South East Technological University,*
Waterford, Ireland
indrakshi.dey@waltoninstitute.ie
0000-0001-9669-6417

*Abstract*—Sleep quality is an important indicator of the efficient cognitive function for high school teachers. Due to the high work stress and multi-tasking expectations, the teachers often face issues with their sleep quality and cognitive function, which has a clearly negative influence on their teaching abilities. In this work, we propose a unique but simple method of deploying Internet of Things (IoT) technology to monitor the sleep quality of high school teachers at Pakistan. Smart watches embedded with pulse rate and SpO$_2$ sensors were used to collect data and categorize the sleep quality as "poor", "fair" or "good". Moreover, we used a psychological tool, Cognitive Assessment Questionnaire (CAQ) for the self-assessment of teachers' cognitive function. The study was conducted over 208 high school teachers from across Pakistan. It has been found that most of the teachers had a poor sleep quality and cognitive function; The link between these two variables indicate that the workload and other factors must be improved for the teachers to ensure their well-being, which will in turn have a positive impact on their teaching quality.

## I. Introduction

Sleep is a fundamental biological function that plays a crucial role in maintaining cognitive health, emotional stability, and overall well-being. Adequate sleep is essential for various cognitive processes such as memory consolidation, decision making, problem-solving, and attention [1]. The importance of sleep is particularly emphasized in high-stress occupations, such as teaching, where mental clarity and emotional regulation are necessary for effective job performance [2]. For high school teachers, whose roles demand constant interaction with students, multitasking, and the management of complex educational tasks, sleep disturbances may significantly impact their cognitive functions and, consequently, their teaching effectiveness [3]. However, despite its importance, sleep quality among teachers often goes overlooked, with many educators experiencing chronic sleep deprivation due to the demands of their profession.

Various methods in Psychology have been developed to monitor the cognitive abilities and sleep patterns of people. Mostly, the psychologists/therapists used well-developed survey tools to collect data about perceptions of individuals regarding their mental health conditions [4]. Although such approaches have been used in clinical psychology for decades, there are drawbacks due to their reliance on self-reported data, which may be influenced by personal bias, memory limitations, or the reluctance of participants to disclose accurate information [5]. Additionally, these methods often lack real-time monitoring capabilities, making it difficult to track fluctuations in cognitive function and sleep quality over time or correlate them with physiological parameters [6]; for example, a person having good sleep quality may feel that they have a poor quality.

Lately, digital technologies have been used to measure the cognitive abilities as well as sleep quality. IoT has been a game changer due to its potential of collecting real-time data using simple, low cost hardware. The technology has been used for a wide range of health and well-being monitoring applications such as chronic disease management [7], elderly care [8], care for disable patients [9], stress and anxiety monitoring [10, 11], fitness tracking [12] and sleep quality assessment [13]. The emerging gadgets such as smart watches have made it possible to collect continuous physiological data from the users, and transmit the same to remote locations.

In this work, we present a unique solution of integrating psychological assessment tool with a real-time IoT system. We used Cognitive Assessment Questionnaire (CAQ) [14] to identify the cognitive function of high school teachers, and also monitored their sleep quality using a smart watch to obtain real time data of their pulse rate and SpO$_2$. We believe that such a system will positively contribute to identify the influence of sleep quality over the cognitive function of teachers, which could in turn be utilized for improving their overall well-being, reducing work-related stress, and designing targeted interventions to enhance their productivity and mental health. To the best of our knowledge, no such system has been proposed in the past.

Rest of this paper has been organized as follows: Section II presents a brief overview of related work. Section III explains the methodology used in this study, along with a brief description of CAQ tool. Section IV presents the results and section V concludes the paper and offers an insight into the future work directions.

## II. Related Work

Research over past several decades has established the fact that sleep quality is a critical factor influencing various aspects of health, including cognitive function [1]. In the context of high school teachers, who face high levels of stress and mental workload, the importance of sleep quality becomes particularly pronounced, as discussed in [14]. Teachers often work long hours, balancing classroom instruction, lesson planning, grading, and extracurricular activities. It has been shown in that the high workload can result in insufficient sleep, leading to disruptions in sleep patterns and a decrease in sleep quality for teachers [15]. Poor sleep can, in turn, negatively affect cognitive functions such as memory, attention, and decision-making, which are essential for effective teaching and classroom management [16]. Research has shown that sleep deprivation and disturbances are linked to cognitive impairments, including reduced concentration, slower reaction times, and difficulty in problem-solving [17]. In particular, teachers are vulnerable to these effects due to the high demands of their profession, which often extends beyond typical working hours. It has been reported in [18] that, the lack of adequate rest can lead to burnout, fatigue, and decreased productivity among teachers.

Various digital technologies have been playing a vital role to measure the sleep quality; among these, IoT and Machine Learning (ML) remain the most prominent. A prototype IoT sleep tracking platform built on physiological and environmental sensors has been developed in [19], with a focus of monitoring and improving sleep quality. A Tele-insomnia framework has been proposed in , for monitoring ECG, Blood Pressure and pulse rate data to determine the sleep quality. IoT has also been integrated with advanced ML techniques to feed in the real-time data and obtain future predictions; for example, a unique approach of integrating hunger games search optimization with deep learning (ASQR-HGSODL) has been proposed in [20], where sleep activity data is collected and sleep quality is predicted.

## III. METHODOLOGY

As mentioned in the previous section, smartwatches and other wearable devices have been extensively utilized to assess sleep quality across various population groups. In this study, we employed smartwatch-based $SpO_2$ and pulse rate sensors to evaluate the sleep quality of high school teachers. The data collected from the smartwatches was then transmitted to a cloud-based dashboard. Sleep quality was classified on the dashboard based on threshold values determined according to the participants' age. We defined three categories: "good," "fair," and "poor." The corresponding threshold values are presented in Table I.

The value from the smart watch was transmitted every 15 minutes, and average was taken. Moreover, another major objective of this research was to integrate IoT with standard Psychology tools for data collection. In this context, we used the Cognitive Assessment Questionnaire (CAQ), comprising 25 items [21]. CAQ is a standardized tool designed to evaluate various cognitive aspects, including memory, attention, reasoning, and problem-solving abilities. It comprises 25 items, each targeting specific cognitive functions to provide a comprehensive assessment. The questionnaire is often used in psychological and behavioral research, particularly in studies related to cognitive load, mental fatigue, and overall cognitive well-being [22]. In the context of this study, CAQ was utilized to assess the cognitive impact of sleep quality among high school teachers, helping to establish correlations between sleep patterns and cognitive performance.

Table 1: Threshold Parameter Values for Measuring Sleep Quality

| Age Group | Parameter | Good | Fair | Poor |
|---|---|---|---|---|
| 25-30 | *Pulse Rate* | 50–65 | 66–75 | >75 or <50 |
|  | *SO_2* | 96–100 | 92–95 | <92 |
| 30-35 | *Pulse Rate* | 55–70 | 71–80 | >80 or <55 |
|  | *SpO_2* | 95–100 | 91–94 | <91 |
| 35-40 | *Pulse Rate* | 58–72 | 73–85 | >85 or <58 |
|  | *SpO_2* | 94–99 | 90–93 | <90 |

We administered the CAQ tool using Google Survey for easy access and data collection. For each question, we offered 5 responses to the participants including "Very Often", "Quite Often", "Occasionally", "Very rarely" and "never". To categorize the cognitive function according to CAQ, we initially assigned the numerical scores to each response like Never = 1, Very rarely = 2, Occasionally = 3, Quite Often = 4, and Very Often = 5. Subsequently, we multiplied each numeric value with the total number of responses which gave us the range: minimum Score = 25 x 1 = 28 (representing lowest cognitive issues), and maximum score = 25 x 5 = 125 (representing highest cognitive issues). Based on this range, the cut-off range defining each sleep category is shown in table II.

Table 2: Cutoff Ranges for Cognitive Function Measurement Using CAQ

| Cognitive Function | Cutoff Range |
|---|---|
| Poor | >= 90 |
| Fair | 60 – 89 |
| Good | < = 59 |

## IV. EXPERIMENTS AND RESULTS

The study was performed on 208 high school teachers of Pakistan for 30 nights, by collecting self-assessment and real-time data. CAQ tool was used to collect data through the Google Survey, and smart watches along with the linked mobile apps were used to collect the pulse rate and $SpO_2$ data. The data was collected each 15 min and was transmitted to the centralized dashboard. The demographics of the study participants has been shown in figure 1. It has been seen that 32% participants were male and 69% were female. Moreover, majority of the participants (53%) were between the age 25 and 30, and majority (38%) had experience of 2 to 3 years.

The survey questionnaire had 25 items. Figure 2 shows the frequency distribution of first 5 questions. It has been seen from the responses that majority of the participants had poor cognitive function at the time of survey; for example, as seen in figure 3-b, majority of the teachers (32%) reported that they quite often forgot why they went from one part of the house to the other, which indicated cognitive function impairment.

Finally, we compared the findings of CAQ tool with those obtained from our IoT solution. Figure 3 shows that both set of findings match with a 90% accuracy. Therefore, it has been found that mostly the teachers had a poor cognitive function along with having poor sleep quality.

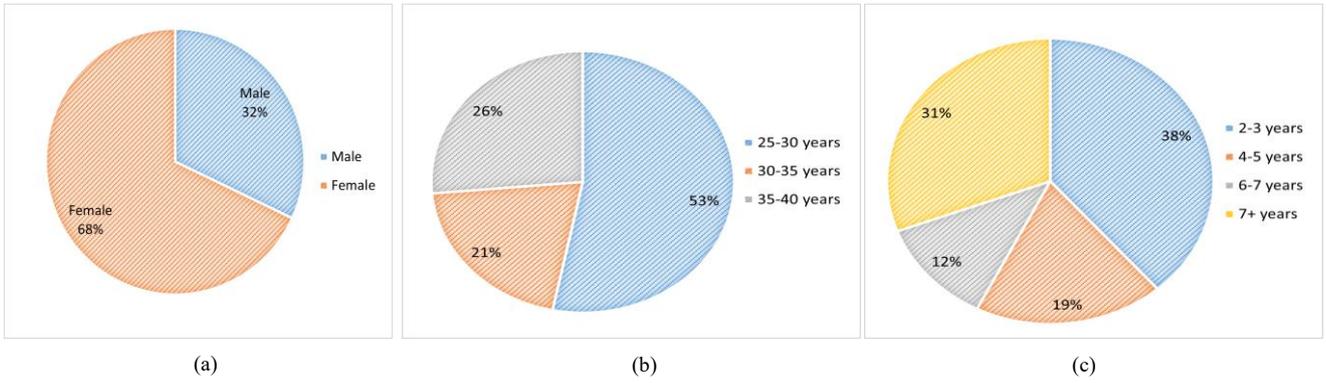

Fig. 1: Demographics of Study Participants: (a) Gender, (b) Age, (c) Experience

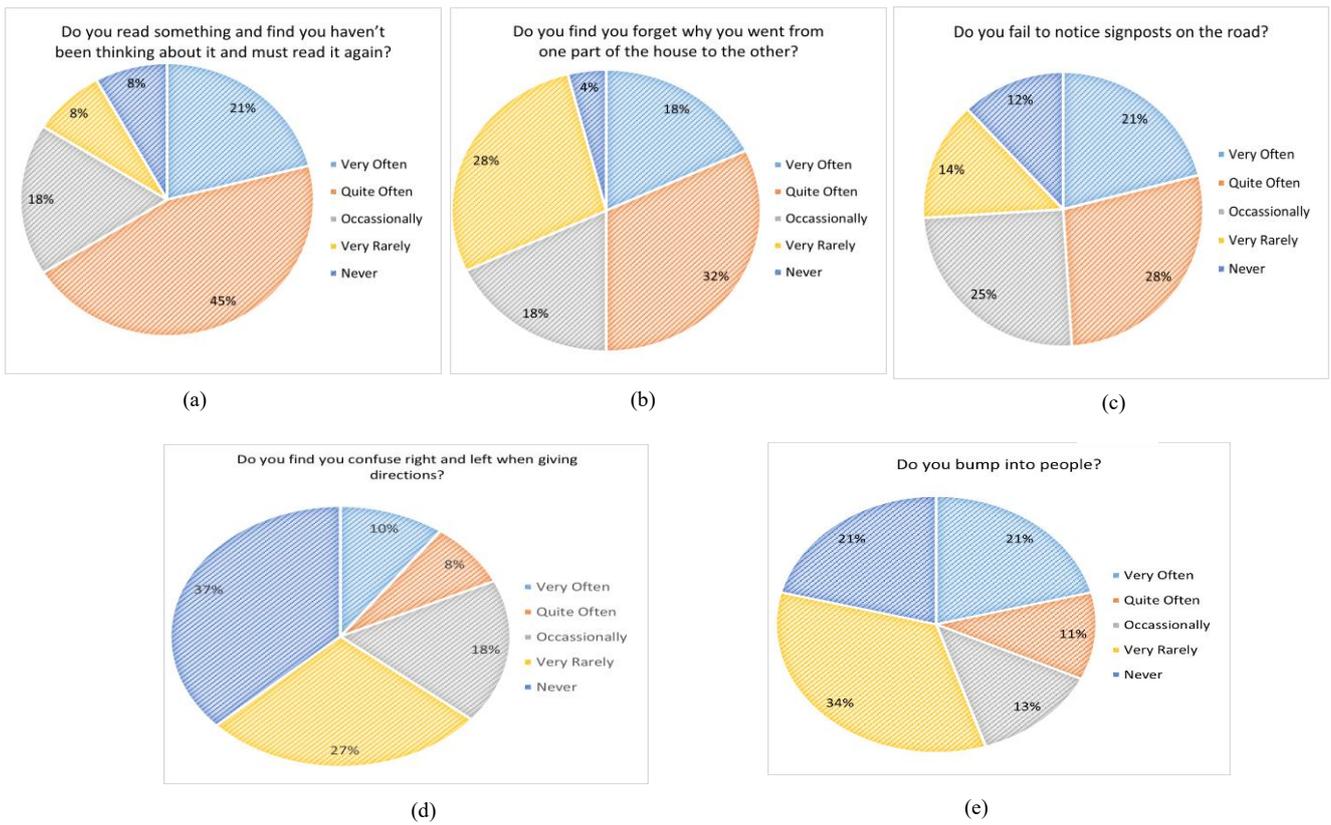

Fig. 2: Frequency Responses of High School Teachers for CAQ Items- First 5 Questions

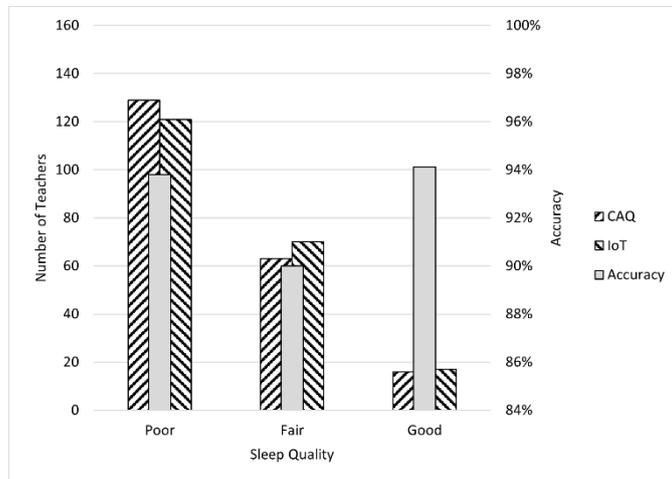

Fig. 3: Comparison of Accuracy for TeachRest and SQS.

## V. Conclusion And Future Work

In this paper, we presented a unique approach of integrating psychology tool CAQ with IoT based sleep quality monitoring solution. The goal of this study was to assess the impact of sleep quality on the cognitive function of high school teachers at Pakistan. It has been found that mostly teachers faced issues with their cognitive function using CAQ tool. On the other hand, using IoT solution, most of the teachers were also observed to having a poor sleep quality. Hence, it has been found that poor sleep quality is linked with poor cognitive function of the high school teachers. Therefore, the present study can be of critical importance for developing effective work stress management strategies for the teachers, and reassess their sleep quality and cognitive function, subsequently.

Future work can extend this study in several directions. First, the research can be expanded to include a larger and more diverse sample of teachers from various countries to improve generalization. Second, additional physiological parameters, such as heart rate variability and movement patterns, can be incorporated into the IoT system for a more comprehensive sleep assessment. Third, integrating machine learning techniques could enhance the accuracy of sleep quality classification and cognitive function predictions.

## Acknowledgment

This contribution is supported by HORIZON-MSCA2022-SE-01-01 project COALESCE under Grant Number 10113073.

## References

[1] K. L. Nelson, J. E. Davis, and C. F. Corbett, "Sleep quality: An evolutionary concept analysis," in Nursing forum, vol. 57, no. 1. Wiley Online Library, 2022, pp. 144–151.

[2] C. J. Travers and C. L. Cooper, "Mental health, job satisfaction and occupational stress among uk teachers," in Managerial, occupational and organizational stress research. Routledge, 2024, pp. 291–308.

[3] P. P. Utami and N. Vioreza, "Teacher work productivity in senior high school." International Journal of Instruction, vol. 14, no. 1, pp. 599–614, 2021.

[4] B. W. Roberts and H. J. Yoon, "Personality psychology," Annual review of psychology, vol. 73, no. 1, pp. 489–516, 2022.

[5] S. Bauhoff, "Self-report bias in estimating cross-sectional and treatment effects," in Encyclopedia of quality of life and well-being research. Springer, 2024, pp. 6277–6279.

[6] T. W. Taris, S. R. Kessler, and E. K. Kelloway, "Strategies addressing the limitations of cross-sectional designs in occupational health psychology: What they are good for (and what not)," pp. 1–5, 2021.

[7] Ullah, M., Narayanan, A., Wolff, A. and Nardelli, P., 2021, September. Smart grid information processes using IoT and big data with cloud and edge computing. In *2021 44th International Convention on Information, Communication and Electronic Technology (MIPRO)* (pp. 956-961). IEEE

[8] A. A. Khan, S. Siddiqui, S. M. Shah, F. Nait-Abdesselam, and I. Dey, "Comparing ANN and SVM algorithms for predicting exercise routines of diabetic patients," in 2021 International Wireless Communications and Mobile Computing (IWCMC). IEEE, 2021, pp. 703–708.

[9] Ullah, M., Hekmatmanesh, A., Savchenko, D., Moioli, R., Nardelli, P., Handroos, H. and Wu, H., 2020, September. Providing facilities in health care via brain-computer interface and Internet of Things. In *2020 43rd International Convention on Information, Communication and Electronic Technology (MIPRO)* (pp. 971-976). IEEE.

[10] S. Siddiqui, A. A. Khan, F. Nait-Abdesselam, and I. Dey, "Anxiety and depression management for elderly using Internet of Things and symphonic melodies," in ICC 2021-IEEE International Conference on Communications. IEEE, 2021, pp. 1–6.

[11] I. Moraiti and A. Drigas, "Measuring the stress of autistic people with the help of a smartwatch, Internet of Things technology," Brazilian Journal of Science, vol. 3, no. 2, pp. 45–56, 2024.

[12] R. Amini Gougeh and Z. Zilic, "Systematic review of IoT-based solutions for user tracking: Towards smarter lifestyle, wellness and health management," Sensors, vol. 24, no. 18, p. 5939, 2024.

[13] S. A. Gamel and F. M. Talaat, "Sleepsmart: an IoT-enabled continual learning algorithm for intelligent sleep enhancement," Neural Computing and Applications, vol. 36, no. 8, pp. 4293–4309, 2024.

[14] P. J. Snyder, C. E. Jackson, R. C. Petersen, A. S. Khachaturian, J. Kaye, M. S. Albert, and S. Weintraub, "Assessment of cognition in mild cognitive impairment: a comparative study," Alzheimer's & Dementia, vol. 7, no. 3, pp. 338–355, 2011.

[15] A. Carrillo-Gonzalez, M. Camargo-Mendoza, and L. C. Cantor-Cutiva, "Relationship between sleep quality and stress with voice functioning among college professors: a systematic review and meta-analysis," Journal of Voice, vol. 35, no. 3, pp. 499–e13, 2021.

[16] A. C. B. de Azevedo, X. d. F. de Medeiros Lopes, J. R. de Lima, C. N. Valenc¸a, D. T. Guedes, and J. C. de Souza, "Relationship between work context and sleep problems of high school teachers," Research, Society and Development, vol. 10, no. 7, pp. e31910716195–e31910716195, 2021.

[17] C. E. Wolff, H. Jarodzka, and H. P. Boshuizen, "Classroom management scripts: A theoretical model contrasting expert and novice teachers' knowledge and awareness of classroom events," Educational Psychology Review, vol. 33, no. 1, pp. 131–148, 2021.

[18] Z. Yang, D. Wang, Y. Fan, Z. Ma, X. Chen, Y. Zhang, and F. Fan, "Relationship between sleep disturbance and burnout among chinese urban teachers: Moderating roles of resilience," Sleep Medicine, vol. 108, pp. 29–37, 2023.

[19] D. J. Jaworski, A. Park, and E. J. Park, "Internet of Things for sleep monitoring," IEEE Instrumentation & Measurement Magazine, vol. 24, no. 2, pp. 30–36, 2021.

[20] B. Venkataramanaiah et al., "Internet of Things assisted sleep quality recognition using hunger games search optimization with deep learning on smart healthcare systems." Journal of Intelligent Systems & Internet of Things, vol. 14, no. 1, 2025.

[21] D. E. Broadbent, P. F. Cooper, P. FitzGerald, and K. R. Parkes, "The Cognitive Failures Questionnaire (cfq) and its correlates," British journal of clinical psychology, vol. 21, no. 1, pp. 1–16, 1982.

[22] W. C. Lim, N. Black, D. Lamping, K. Rowan, and N. Mays, "Conceptualizing and measuring health-related quality of life in critical care," Journal of critical care, vol. 31, no. 1, pp. 183–193, 2016.